\documentclass[sigconf,screen]{acmart}
\usepackage{booktabs}
\usepackage{color}
\usepackage{balance}

\def\BibTeX{{\rm B\kern-.05em{\sc i\kern-.025em b}\kern-.08emT\kern-.1667em\lower.7ex\hbox{E}\kern-.125emX}}

\copyrightyear{2020}
\acmYear{2020}
\setcopyright{acmlicensed}\acmConference[UMAP '20]{Proceedings of the 28th ACM Conference on User Modeling, Adaptation and Personalization}{July 14--17, 2020}{Genoa, Italy}
\acmBooktitle{Proceedings of the 28th ACM Conference on User Modeling, Adaptation and Personalization (UMAP '20), July 14--17, 2020, Genoa, Italy}
\acmPrice{15.00}
\acmDOI{10.1145/3340631.3394864}
\acmISBN{978-1-4503-6861-2/20/07}

\begin{document}
\title{Beyond Optimizing for Clicks: Incorporating Editorial Values in News Recommendation}


\author{Feng Lu}
\authornote{Now at Bol.com}
\affiliation{%
  \institution{FD Mediagroep}
  \city{Amsterdam}
  \country{The Netherlands}
}
\email{publicflu@gmail.com}

\author{Anca Dumitrache}
\authornote{Now at Talpa Network}
\affiliation{%
  \institution{FD Mediagroep}
  \city{Amsterdam}
  \country{The Netherlands}
}
\email{anca.dmtrch@gmail.com}

\author{David Graus}
\authornote{Now at Randstad Groep Nederland}
\affiliation{%
  \institution{FD Mediagroep}
  \city{Amsterdam}
  \country{The Netherlands}
}
\email{dpgraus@gmail.com}


\begin{abstract}
  With the uptake of algorithmic personalization in the news domain, news organizations increasingly trust automated systems with previously considered editorial responsibilities, e.g., prioritizing news to readers. 
  In this paper we study an automated news recommender system in the context of a news organization's editorial values. 
  
  We conduct and present two online studies with a news recommender system, which span one and a half months and involve over 1,200 users.
  In our first study we explore how our news recommender steers reading behavior in the context of editorial values such as serendipity, dynamism, diversity, and coverage. 
  Next, we present an intervention study where we extend our news recommender to steer our readers to more dynamic reading behavior. 
  
  We find that 
  (i) our recommender system yields more diverse reading behavior and yields a higher coverage of articles compared to non-personalized editorial rankings, and 
  (ii) we can successfully incorporate dynamism in our recommender system as a re-ranking method, effectively steering our readers to more dynamic articles without hurting our recommender system's accuracy. 
\end{abstract}



\maketitle

\section{Introduction}
\label{section:intro}

The news media have undergone a substantial transformation with the advent of algorithmic personalization and recommendation~\cite{newsrecjannach}, which is increasingly employed as means to improve access to the increasing amounts of news sources and articles online~\cite{sappelli2018smartjournalism, sappelli2018smartradio}.
However, with this transformation, parts of what are traditionally considered editorial responsibilities, are transferred to algorithms and automated systems~\cite{carlson2018}. 
A complicating factor in this shift of power is that traditionally, recommender systems learn from and optimize for historic user behavior, i.e., clicks~\cite{Oard98implicitfeedback}. 
Research has since identified several `usefulness' metrics that aim to provide insights beyond accuracy, which can be of importance in assessing a recommender system's quality, e.g., serendipity and coverage~\cite{10.1145/1864708.1864761}. 

News recommendation differs from many traditional recommendation domains such as e-commerce or entertainment, in that news organizations 
both have a clear responsibility towards society~\cite{helberger2019}, and 
also typically uphold their own journalistic or editorial values, as a framework for their journalism. 
For these reasons, we argue that in the news domain we need to move beyond only addressing users' perception~\cite{thurman2019,bodo2019}, and also consider providers (news organizations) as stakeholders.
As the role and purpose of algorithmic personalization may differ between news organizations~\cite{bodo2019selling}, strictly performance-driven optimization may not be a suitable strategy, turning attention to more fine-grained `usefulness' metrics such as diversity or serendipity. 

In this paper we study a news organization's journalistic values that are considered of importance in the context of algorithmic personalization.
These values are: 
(i) the ability to \emph{surprise} readers, 
(ii) providing \emph{timely and fresh news},
(iii) yielding more \emph{diverse} reading behavior, and 
(iv) increasing item \emph{coverage}.
We set out to explore how our news recommender can effectively incorporate these values algorithmically. 
We do so by performing two online user studies.

First, we aim to answer the following research question: \textbf{RQ1:} \emph{``Does our recommender system effectively steer users to useful recommendations?''}
We answer this question by analyzing the news recommender of ``Het Financieele Dagblad'' (FD)\footnote{\url{https://fd.nl/}}, a Dutch newspaper in the financial economic domain, on four usefulness metrics: diversity, coverage, serendipity, and dynamism.
We find that our news recommender presents our users with more diverse and serendipitous articles, compared to manually curated lists of articles. 
More importantly, we see these recommendations successfully steer our users towards more diverse consumption, with an increased item coverage from the provider's perspective compared to manually curated news.

Next, we answer \textbf{RQ2:} \emph{``Can we effectively adjust our news recommender to steer our readers towards more dynamic reading behavior, without loss of accuracy?''}
We do so by studying dynamism in an intervention study, which is identified as an important editorial value in the context of algorithmic personalization at FD.
We implement dynamism in our news recommender as a re-ranking strategy, and expose users to different treatments to measure its impact. 
We find we can effectively incorporate dynamism without loss of accuracy, while successfully steering our users to more dynamic reading behavior.

\section{Related Work}
The presented work touches on several areas, from the role and impact of recommender systems in the news domain, the technical challenges and aspects of news recommendations, and recommender systems metrics that aim to move beyond optimizing for clicks. 

\subsection{Algorithmic personalization in news}
The rise of algorithmic personalization calls into question how editorial and algorithmic responsibilities relate~\cite{carlson2018}.
On the one hand, audiences believe algorithmic recommendation is a better way to get news than editorial curation, but the strength of this belief varies by demographics~\cite{thurman2019}. 
Similarly, the perceived value of news recommendations depend on users' expectations, which differ by demographics too~\cite{bodo2019}. 

On the other hand, recommenders play an important role and have a responsibility in a news organization's (democratic) mission~\cite{helberger2019}.
There is a wide variety of ``perceived'' roles for algorithmic personalization in the newsroom, from simply selling (more) articles and subscriptions (or optimizing similar business metrics~\cite{businessvaluejannach}), to serving under-served audiences~\cite{bodo2019selling}.
At the same time, newsrooms are aware of the importance of editorial values in algorithmic design~\cite{bastian2019}, e.g., for transparency~\cite{sullivan2019reading}.
But the exact role of editorial values in algorithms often stays unclear~\cite{vandrunen2019}. 
We set out to address this by explicitly incorporating editorial values into our news recommender system.

\subsection{News recommender systems}
The news domain has the constraint of continuous item cold start: once an article is published, the typically limited shelf life means it is important to recommend it as soon as possible.
These types of constraints means that collaborative filtering approaches are not the method of choice in news recommendation~\cite{newsrecjannach}, as they commonly rely on having to acquire enough user signal to effectively recommend an item, turning attention to content-based methods.

In a similar scenario, \citet{odijkschuth2017} employ an online learning to rank-powered content-based approach for news recommendation, using features that relate to the user, the article, and the intersection of both. 
In related domains we see similar methodologies, e.g., in e-commerce~\cite{freno2017practical}, and in targeted advertising~\cite{he2014practical} too, learning to rank-based methods, trained on implicit feedback, are commonly used. 
We provide more details on how the aforementioned work relates to FD's news recommender in Section~\ref{section:system-design}.

\subsection{Beyond accuracy: `usefulness' in recommender system evaluation}
When business values, editorial values, and algorithms coincide, the question naturally arises what to measure, and what to optimize your algorithms for. 
Recommender systems literature contains a rich body of work studying evaluation in general, which is a non-trivial problem~\cite{shani2011evaluating}. 
Different metrics that aim to move beyond ``simple'' accuracy~\cite{kaminskas2017-beyondaccuracy} have been proposed, such as diversity~\cite{vargas2011,bradley2001improving,lathia2010temporal,anelli2017} and novelty~\cite{mcnee2006being}, or serendipity and coverage~\cite{10.1145/1864708.1864761}. 
These additional metrics aim to evaluate the quality of recommender systems in different dimensions, that move beyond simple optimization for clicks, and aim to capture aspects of recommender system's usefulness. 

In our work we employ some of the aforementioned metrics to measure the recommender system's output, and to understand the impact of recommendations on our users' reading behavior.

\begin{figure*}[h!]
    \centering
    \includegraphics[width=.8\textwidth]{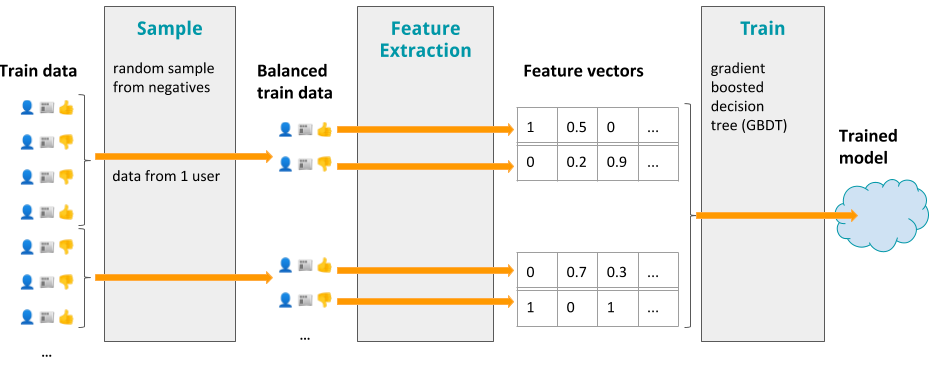}
    \caption{Daily recommender system train pipeline.}
    \label{fig:train}
\end{figure*}

\section{News Recommender System Design}
\label{section:system-design}

This section describes the architecture and features of the recommender system we studied. Since recommending freshly published news is a cold start problem by nature, we employ a content-based model. An overview of the recommender model training process is shown in Figure~\ref{fig:train}.

\subsection{Data}
The data used for the training, validation and evaluation of the model consists of implicit feedback collected from users logged into the news website -- article clicks are labeled as positive, and article links that were seen, but not clicked by the user are labeled as negative. Similarly to other news datasets~\cite{kille2013plista}, we observed an imbalance in the label distribution, with the negative labels outnumbering the positives. In order not to overfit the model on negative data, we performed random negative sampling to get an equal ratio of positive and negative labels.

\subsection{Features}
Inspired by similar work in news recommendations~\cite{odijkschuth2017}, we experimented with a set of 60 features, representing the article and user, in addition to hybrid features that measure the compatibility between the user and article. 
The \textbf{article features} consist of metadata about the article and extracted content features, resulting in a heterogeneous mix of feature types:

\begin{itemize}
    \item \textit{categorical article features:} tags, authors, website section;
    \item \textit{embedded article features:} average word embeddings (taken from pre-trained fastText Dutch language vectors\footnote{\url{https://fasttext.cc/docs/en/crawl-vectors.html}}) of all the words in the article;
    \item \textit{temporal features:} hour of publication, day of the week;
    \item \textit{stylometry features:} hapax legomena, dis legomena; 
    \item \textit{length-based article features:} number of words, number of sentences, number of paragraphs, article length.
\end{itemize}

Out of a concern for user privacy as well as model bias, the \textbf{user features} do not use user demographics, but instead focus on aggregated user reading behavior, such as the most read authors and tags, the average article length, and the average word embeddings of all articles read. 
Finally, the \textbf{user-article features} contain both overlap features (e.g., number of tags in common between the article and user data), as well as comparison features (e.g. article length compared to the average length read by the user).

\subsection{Model}
Based on previous work by~\citet{he2014practical}, we experimented with a \textbf{gradient boosted decision tree (GBDT)} architecture, with a logistic regression final layer. An experiment on offline data found that the simple GBDT performed best in our scenario. 
We use the model's confidence scores per article to generate a ranked list of articles for each user. 
We train a new model nightly, using user interaction data from the previous 7 days. 
At prediction time, we rank a candidate set of articles from the previous 7 days. 
Ranked lists of articles for each user are generated at pre-selected times and cached in a database. 
Each morning, a batch job ranks a list of articles for all users to account for the morning news articles, and then prediction jobs are scheduled to occur hourly, in addition to being triggered after users visit the website, so that newly published articles will appear in the ranked list for the user's next visit. 

The model architecture, optimal hyperparameters of the model, and the optimal number of 7 days for training were tuned in an offline experiment, that used interaction data from the news website over the span of one month. 
The model was implemented using the XGBoost~\cite{chen2016xgboost} library. 
A full list of the features, as well as the full model's optimal hyperparameter values are provided in the supplementary material~\cite{lu_feng_2020_3758172}.

\section{User Study 1: Usefulness Analysis}
\label{section:study_1}

In this section we present our first user study, where we examine our news recommender's effect on reading behavior. 
We first establish the overall performance of our recommender system in an offline evaluation in Section~\ref{section:alpha_test}. 
Then, we answer \textbf{RQ1} by measuring several aspects of recommendation `usefulness,' and their effects on reading behavior. 
More specifically, we introduce and describe four usefulness metrics, which we measure and compare between recommendations and manually curated articles in Section~\ref{section:usefulness}.
Finally, in Section~\ref{section:historic} we compare reading behavior from recommendations to manually curated articles, by comparing behavior before and after introduction of the news recommender, to measure the extent in which our news recommender steers reading behavior.

\subsection{Presentation of recommendations}
\label{sec:pres}

Clicks for recommended articles are collected from three different sections of the website:

\begin{itemize}
    \item \textsc{MyNewsWidget} (MNWidget) is a widget shown in the top-right corner of the front page (``above the fold'') that shows the top 5 recommended articles published in the last 24 hours. 
    The section is meant to allow readers to catch up with the latest news that is relevant to them.
    \item \textsc{MissedLastWeek} (MissedLW) is a section on the front page (``below the fold'') that shows the top 5 of recommended articles published in the last seven days, but that are older than 24 hours. 
    The section is meant to highlight interesting articles that the readers might have missed on the front page in the last week.
    \item \textsc{MyNewsPage} (MNPage) is a separate page that lists all recommended articles. 
\end{itemize}

\begin{figure}[t]
   \centering
   \includegraphics[width=\linewidth]{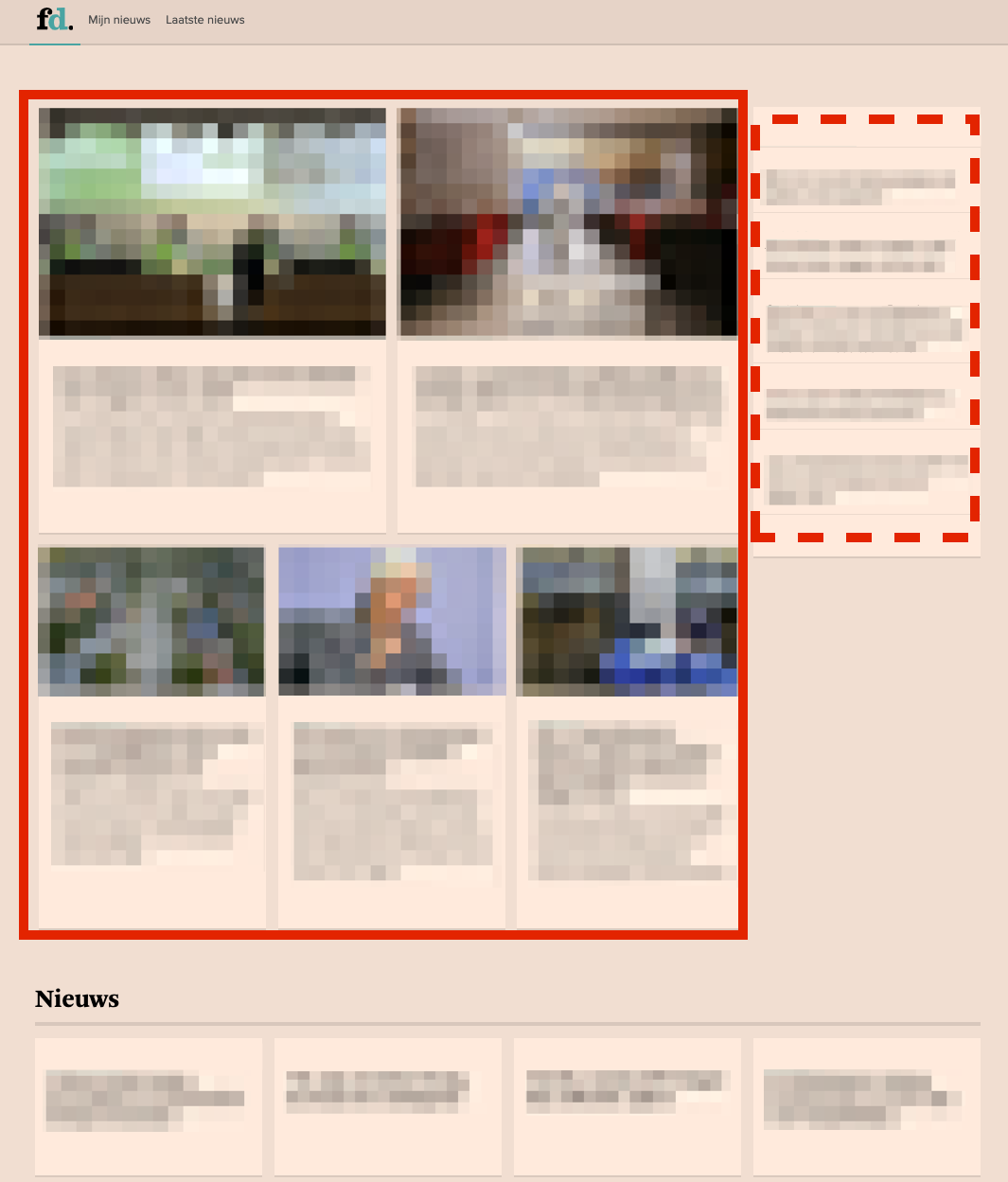}
   \caption{Front page layout. The ``highlighted'' articles are shown in the solid box (left), the widget in the dashed box.}
   \label{img:high5}
 \end{figure}

The \textsc{MyNewsWidget} and \textsc{MissedLastWeek} sections are shown on the front page as a widget and horizontal list of items respectively, and hence only permit to display the top 5 articles of the recommended article lists, whereas the \textsc{MyNewsPage} is a dedicated page, which shows all articles available to be ranked in a vertical list (for an illustration, see Figure~\ref{img:high5}, where the \textsc{MyNewsWidget} is shown in the top right dashed box, and the \textsc{MissedLastWeek} is rendered like the bottom \emph{``Nieuws''} ribbon). Per section, articles are ranked based on the confidence score of the \textsc{RecSys} ranker, as described in Section~\ref{section:system-design}.

In addition, any article on the front page also receives a \textsc{RecommendedLabel}, if the confidence score of the model for that article is $ >= 0.5$. Clicks on items with this label also count as recommended artile clicks.

As a baseline for our experiment, we consider the \textsc{manual} ranker - an editorially curated non-personalized top 5 of highlighted articles. These appear on the website in a grid at the very top (above the fold) of FD's frontpage, spanning the full width of the main content column. Figure~\ref{img:high5} shows an illustration of these ``highlighted'' articles, in the left grid with 2 wide and 3 narrow articles.

\subsection{Accuracy}
\label{section:alpha_test}
We answer \textbf{RQ1} by conducting an online test, spanning one month where we log user interactions of a group of 115 users. 
During this test, we measure and compare our readers' reading behavior on manually curated and algorithmically personalized lists of articles, by comparing user clicks on articles from the \textsc{RecSys} ranker and the \textsc{manual} ranker.
To get a sense of general system performance, we first report our system's accuracy in precision, recall, and NDCG, using an offline evaluation methodology.

\subsubsection{Offline evaluation}
We evaluate our recommender system as follows. 
First, we capture our users' clicked articles per day, which we consider positive samples, and all articles that were displayed but not clicked, which we consider negative samples. 
We then employ each recommendation model that was trained on clicks up to the day prior to the collected positive and negative samples, to simulate how the clicked articles would have been ranked in the candidate lists. 

We measure the NDCG scores on the simulated recommendations for each user, and report the averaged NDCG scores for all users for all days. 
In addition, since our front page highlighted section contains 5 articles, for easier comparison, we report precision and recall at the top 5 (P@5 and R@5) and top 10 (P@10 and R@10) recommendations.

\paragraph{Results}
Table \ref{tab:offline_performance} shows the performance of the recommender system. 
The NDCG score (0.71) tells us that users' clicked articles ranked relatively high in our recommendations. 
Recall scores (0.55, 0.74) show us that over half of the clicked articles rank among the top 10 recommendations. 
Overall, without having a baseline to compare against, we believe the metrics point to an adequate ability of the recommender system to rank read articles highly.
We revisit how these accuracy metrics compare to online performance in Section~\ref{section:online_accuracy}.

\begin{table}[t]
    \caption{Offline evaluation performance.}
    \label{tab:offline_performance}
    \begin{tabular}{@{}llllll@{}}
    \toprule
     & NDCG & R@5 & P@5 & R@10 & P@10 \\ \midrule
    RecSys & 0.71 & 0.55 & 0.34 & 0.74 & 0.25 \\ \bottomrule
    \end{tabular}
\end{table}

\subsection{Usefulness}
\label{section:usefulness}

The core of this study revolves not around accuracy, but `usefulness' of our recommendations.
To answer \textbf{RQ1}, we compare each user's top 5 recommendations (\textsc{RecSys}) to the front page 5 highlighted articles (\textsc{manual}) on diversity, dynamism, serendipity, and coverage, as introduced in Section~\ref{section:intro}.

\subsubsection{Usefulness 1: Diversity}
Diversity is usually considered as the inverse of similarity~\cite{ricci2011introduction}, which refers to recommending a diverse set of items to users so as to help them discover unexpected and surprising items more effectively~\cite{mcnee2006being}.

\paragraph{Method}
We employ the commonly used \emph{intra-list diversity}~\cite{bradley2001improving,kaminskas2017-beyondaccuracy} of a list of articles as follows:

\begin{equation}
    Div(R) = \frac{\sum^{n}_{i=1}\sum^{n}_{j=i}(1-Sim(c_i,c_j))}{n\cdot(n-1)/2},
\end{equation}
where $c_1$,\ldots,$c_n$ are items in a set of recommendation list, $R$ is the list of recommendations, and $Sim()$ a similarity metric.

We measure similarity between articles using different article attributes: 
\emph{author(s)}, \emph{tags}, \emph{sections}, and \emph{word embeddings}. 
The `authors' and `tags' attributes were found to be two of the most important user-article features in our model, which represent the article's author(s), and a list of assigned tags (keywords) from a predefined list. 
The `sections' attribute represents a (broad) categorization of the article, taken from a pre-defined list that editors input in the CMS.
`Word embeddings' represents the article content as an averaged word embedding vector, as explained in Section~\ref{section:system-design}. 

We use different similarity metrics for different attributes, for discrete attributes (`section', `tags', and `authors') we use Jaccard Index. We verified our findings with different diversity metrics, e.g., Gini coefficient and Shannon Entropy \cite{shani2011evaluating}, which showed consistent results. 
The `word embeddings' are dense vectors, so we employ the commonly used cosine similarity (normalized by the maximal score). 

To compare diversity between our \textsc{manual} and \textsc{RecSys} rankings, we first need to temporally align both. 
The \textsc{manual} lists are updated irregularly at different moments during the day, and our \textsc{RecSys} updates in regular intervals at a higher frequency (hourly).
For this reason, we align both sources' updates of rankings at the timestamps of when \textsc{manual} changes. 
We find 377 \textsc{manual} lists (i.e., updates of rankings) during the span of our test (on average $\sim$12 per day). 
Our alignment procedure hence yields $377\cdot115$ \textsc{RecSys} rankings.

For each ranking list, we calculate the diversity scores for all selected article attributes.
Then we average all lists' scores separately for \textsc{manual} and \textsc{RecSys}.
For each attribute diversity comparison, we perform student's t-test on the two score sets, effectively assessing whether the average scores differ significantly between our two treatments (\textsc{manual} and \textsc{RecSys}).
We employ the same statistical testing methodology for all other experiments in the rest of the paper. 

\paragraph{Results}
Table \ref{tab:divr} shows the comparison on the intra-list diversity of the article attributes described above. 
From the table, we see that the top 5 recommendations are more diverse than the highlighted articles in terms of `section' and `word embeddings.' 
Since there are many hundreds of unique tags, they typically exhibit a low overlap between articles, explaining the relatively high diversity in both sources.
In terms of `authors,' the manually curated highlighted articles show higher diversity than recommended articles. 
This can be explained by our finding that `authors' and `tags' are two of the most important user-article features in our models, which means recommendations will tend to be personalized more strongly towards `authors' and `tags' that are similar to our users' reading history.

\begin{table}[t]
    \caption{Diversity per article attribute. * indicates statistically significant difference with $p$\textless 0.05.}
    \label{tab:divr}
    \begin{tabular}{@{}rcc@{}}
    \toprule
    Attribute & \textsc{manual} & \textsc{RecSys} \\ \midrule
    Section & 0.6045 & \textbf{0.7370*} \\
    Tags & 0.9576 & \textbf{0.9619*} \\
    Authors & \textbf{0.9291*} & 0.8724 \\
    Word Embeddings & 0.1152 & \textbf{0.1357*} \\ \bottomrule
    \end{tabular}
\end{table}

\subsubsection{Usefulness 2: Dynamism}
Providing dynamic rankings and delivering timely and fresh recommendations is of central importance to a news recommender.  
We measure dynamism by measuring \emph{inter-list diversity}, or how much an article lists changes between two updates~\cite{lathia2010temporal}.

More specifically, we measure the percentage difference between two consecutive rankings as follows:

\begin{equation}
    diversity(L^1, L^2, N) = \frac{|L^2 \setminus L^1|}{N}.
\end{equation}

Here $L^1$ and $L^2$ are two (consecutive) recommendation lists, and $N$ is the length of the recommendation lists.

\paragraph{Method}
We compute and compare dynamism scores for the manually curated front page articles (377 lists) to two versions of lists of recommendation:
(i) \emph{aligned} recommendations, where we take the recommendation list at the timestamp of an updated \textsc{manual} list as described above (yields ($377\cdot115$ lists), and 
(ii) \emph{all} changes of recommendations, where we consider each update of a recommendation list, irrespective of the \textsc{manual} updates (38,343 unique lists). 

\paragraph{Results}
In Table~\ref{tab:dynamism}, we see that the manually curated articles are more dynamic than the aligned recommendations ($0.32>0.16$). 
This may be explained by the fact that recommendation lists only change when (i) articles are published, or (ii) users read articles, and are otherwise static. Whereas, editors regularly change the articles shown on the front page. 
However, when we look at all the list changes, the recommendations are instead shown to be more dynamic ($0.42>0.32$). 
This shows that our top 5 recommendation lists might not change frequently (e.g., seldom changing per hour), but once they change, they introduce more new items to the list, and hence are more dynamic.

\begin{table}[t]
    \caption{Dynamism. * indicates statistically significant difference compared to \textsc{manual}, with $p$\textless 0.05}
    \label{tab:dynamism}
    \begin{tabular}{@{}rcc@{}}
    \toprule
    \textsc{manual} & \textsc{RecSys} (aligned) & \textsc{RecSys} (all) \\ \midrule
    0.3218 & 0.1628* & \textbf{0.4167}* \\ \bottomrule
    \end{tabular}
\end{table}

\subsubsection{Usefulness 3: Serendipity}
As described by~\citet{10.1145/1864708.1864761}, serendipity is concerned with \emph{``in how far recommendations may positively surprise users''}~\cite{10.1145/1864708.1864761}.
We model serendipity similarly to~\citet{10.1145/1864708.1864761}, and consider it a balance between \emph{usefulness}, as represented by the recommender system's confidence score, and \emph{unexpectedness}, which we model as an article's dissimilarity to a reader's ``expected'' (i.e., historic) reading behavior. 

\paragraph{Method}
We aggregate the reader's reading history, and compare its similarity to each ranked article in a list, which we aggregate and average. 
We use different similarity metrics for different attributes:
for discrete attributes (authors, tags, sections) we employ the Gini coefficient, 
for our continuous attribute (word embeddings), we employ cosine similarity.

In the former case, we represent the user's reading history as the aggregated set of all discrete items (e.g., tags, authors, sections) of the user's past seven days' reading history. 
In the latter case we represent their history as the averaged word embedding from their past seven days' reading history.

\paragraph{Results}
Table~\ref{tab:divrh} shows the results for the averaged serendipity comparison between \textsc{manual} and \textsc{RecSys}. 
We find no significant difference in section serendipity, but we do find \textsc{manual} yields more serendipitous rankings in `tags' and `authors' than \textsc{RecSys}.
The latter is more serendipitous in `word embeddings.'
This is expected, as `tags' and `authors' are two important user-article features, which means recommendations will be steered towards more similar tags and authors (see also our observations with Diversity, above).

\begin{table}[t]
    \caption{Serendipity per attribute. * indicates statistically significant difference with $p$\textless 0.05.}
    \label{tab:divrh}
    \begin{tabular}{@{}rcc@{}}
    \toprule
    Attribute & \textsc{manual} & \textsc{RecSys} \\ \midrule
    Section & 0.4465 & 0.4381 \\
    Tags & \textbf{0.1758*} & 0.2060 \\
    Authors & \textbf{0.2187*} & 0.2754 \\
    Word Embeddings & 0.7009 & \textbf{0.7680*} \\ \bottomrule
    \end{tabular}
\end{table}

\subsubsection{Usefulness 4: Coverage}
Another aspect that is important for a news recommender is how much of the archive it is able to serve to its readers.
One strength of algorithmic personalization is it is tailored to each reader, meaning more specific content can be served to specific audiences, effectively exposing each article to its potential audience~\cite{wasilewski2018}. 

We model coverage as the percentage of daily published articles that are served in a list. 
We compute coverage scores for \textsc{RecSys} per user, which we aggregate across all users, and compare these to the coverage of the non-personalized manually curated front page (\textsc{manual}). 

\paragraph{Results}
Table~\ref{tab:Coverr} shows the results. 
Of the 70 articles that are published daily on average, around one third (30\%) are featured on the front page~\cite{sappelli2018smartjournalism}. 
When we look at coverage per user, the top 5 recommendations for each user only cover around 12\% of all publications.
However, looking at the recommendation coverage aggregated across all users, we find that the top 5 recommendations cover 77\% of all publications.

\begin{table}[t]
    \caption{Coverage. * indicates statistically significant difference compared to \textsc{manual} with $p$\textless 0.05.}
    \label{tab:Coverr}
    \begin{tabular}{@{}rcc@{}}
    \toprule
    \textsc{manual} & \textsc{RecSys} (per user) & \textsc{RecSys} (all users) \\ \midrule
     0.2995 & 0.1167* & \textbf{0.7748*} \\ \bottomrule
    \end{tabular}
\end{table}

This tells us that per user the recommendations may provide a narrow set of articles, since the recommender system aims to be as personalized as possible. 
However, across all users, with each user having distinct preferences, the overall coverage of recommended articles is much higher than the manual selection (which is tailored to everyone).

\begin{table}[t]
    \caption{Reading behavior compared between before (July) and after (August) introduction of the news recommender. * indicates statistically significant difference with $p$\textless 0.05.}
    \label{tab:divc}
    \begin{tabular}{@{}rcc@{}}
    \toprule
    Attribute & July & August \\ \midrule
    Div$_{Section}$ & 0.4840 & \textbf{0.5139*} \\
    Div$_{Tags}$ & 0.6216 & \textbf{0.6658*} \\
    Div$_{Authors}$ & 0.5827 & \textbf{0.6229*} \\
    Div$_{Word Embeddings}$ & \textbf{0.2565*} & 0.2463 \\ 
    Coverage & 0.7607 & \textbf{0.8231*} \\
    \bottomrule
    \end{tabular}
\end{table}

\subsection{Effect on reading behavior}
\label{section:historic}
Finally, we study our test users’ reading behavior before and after introduction of the news recommender, to understand whether the recommender system successfully steers our readers to more useful reading behavior.

We collect all article clicks of our test users in the month prior to running the user study (i.e., before the news recommender was deployed), and collect their clicks during the user study (i.e., which includes clicks on recommendations).
We compute the usefulness metrics over the collected articles, and compare them between July and August, to understand how the reading behavior differed between the two months. 

Table \ref{tab:divc} shows the results for Diversity and Coverage. 
The table shows that when the recommender system is introduced, users' daily article clicks are more diverse on every attribute except for word embeddings, which suggests our recommender system effectively steers users towards more diverse reading.
In addition, the coverage (aggregated over all users) substantially increases with the introduction of the recommender system, suggesting it successfully finds the target audiences for the daily published articles~\cite{wasilewski2018}.

\subsection{Summary}
In our first user study, we find that the recommender system successfully ranks historically clicked articles highly. 
In addition, our recommender system presents readers with more diverse articles in terms of topics and content than manually curated articles. 
The recommender system yields less frequent but more thorough changes in rankings. 
We find the recommender system surprises readers less on tags and authors, but more in terms of content (word embeddings) than manually curated lists. 
Finally, while the recommender system yields a lower coverage at the individual level, from the provider's perspective, coverage increases substantially. 
Moreover, by comparing reading behavior before and during the introduction of the recommender system, we show it successfully steers readers towards more diverse reading with higher item coverage.

\section{User Study 2: Editorial value-steered recommendation}
\label{section:beta_test}
Having identified dynamism as an important editorial value for algorithmic personalization, we set out to answer \textbf{RQ2}: \emph{``Can we effectively adjust our news recommender to steer our readers towards more dynamic reading behavior?''} 
We do so by running an A/B test with recommendations biased towards higher dynamism. 

In this section, we first describe why and how we incorporate dynamism in our news recommender.
Next, we establish whether our treatment has the expected result on the recommender system. 
Finally, we compare our recommender system's accuracy with and without additional dynamism, to establish whether we can successfully steer reading behavior without loss of accuracy.

\subsection{Editorial Values}
\label{section:editorial-values}
Our news organization participated in a study by~\citet{bastian2019} in which journalistic values emerged that are considered important in the context of news recommendation. 
We followed this study up with our own interviews and meetings between FD's developers, data scientists, and journalists, which resulted in the identification of two values that were both considered important in the context of algorithmic news personalization, and feasible in technical implementation: 
(i) the recommender system should always yield \emph{timely and fresh} content, which we model as \emph{dynamism}, and 
(ii) the recommender system should be able to \emph{surprise} readers, which can be modeled as \emph{serendipity}. 

Because we want to avoid exposing readers to sub-optimal rankings, and are constrained by technical requirements (see also Section~\ref{section:limitations}), we limited our intervention to incorporating \emph{dynamism}, which was determined both a feasible metric to implement and a low-risk adjustment of the recommender system's output.

The A/B test we conduct in our second user study hence contains the following two treatments: 
(i) the original recommender system (\textsc{Baseline}), and 
(ii) the recommender system steered towards more dynamic recommendations (\textsc{dynamism}).

\subsection{Dynamism}
Our dynamism computation boils down to re-ranking recommendations by incorporating a measure of the article \emph{recency}, to rank more recently published articles higher.
We expect this to increase dynamism as defined in User Study 1 (intra-list diversity), as lists will change more when new articles are published. 
We compute dynamism as follows:

\begin{equation} \label{eq:dynamism}
    Dyn(a) = 1 - \dfrac{1}{1 + \log[1 + (t(a) - t(start)) / 3600]},
\end{equation}
where $t(a)$ is the (publication) timestamp of article $a$, and $t(start)$ is the timestamp of the start of the online user study.

We incorporate dynamism into our recommender system with a linear re-ranking method, where we combine the metrics with the recommender system's confidence score as follows:

\begin{equation} \label{eq:usefulness}
    \lambda S(u,a) + (1-\lambda) Dyn(u,a),
\end{equation}
where $S(u,a)$ is the original model confidence score for article $a$ and user $u$, $Dyn(u,a)$ represents our dynamism computation, explained in more detail in equation~\ref{eq:dynamism}, and $\lambda$ is the ratio coefficient controlling the balance between dynamism and the original confidence score.
We set $\lambda=0.5$, which we empirically determine to be optimal on the same offline data used to tune the model in Section~\ref{section:system-design}.

\subsection{Online Test}
\label{section:online-test}
We ran our online A/B test as part of a bigger online test for FD.nl for a period of two weeks (November 25 to December 4, 2019), to a group of 1,108 readers.
Each reader was randomly assigned to one of our two treatments: \textsc{Baseline} or \textsc{Dynamism}. 
Our readers opted in for participating in the online test, and we only approached long-term readers for participation. In this test, we display recommended articles in the same three sections described in Section~\ref{sec:pres}.

Per section, articles are ranked based on the test treatment, either with the news recommender's confidence score $S(u,a)$ or the combined scores given by Equation~\ref{eq:usefulness}.

\subsection{Treatment effectiveness}
To study whether our dynamism treatments yields the expected effect, we measure the usefulness metrics presented in Section~\ref{section:usefulness} on the aggregated rankings per treatment, which we also aggregate and average across the different presentation sections. 

\begin{table}[t]
    \caption{Usefulness metrics of the treatments in User Study 2. * indicates statistically significant difference with $p$\textless 0.05}
    \label{tab:beta_treatments}
    \begin{tabular}{@{}lllll@{}}
    \toprule
    $ \frac{metric \rightarrow}{\downarrow treatment}$ & Dyn & Ser & Cov & Div \\ \midrule
    \textsc{Baseline}    & 0.9460 & 0.6276 & 0.3318 & 0.0851 \\
    \textsc{Dynamism} & \textbf{0.9799*} & \textbf{0.6497*} & 0.3205 & \textbf{0.0921*} \\ \bottomrule
    \end{tabular}
\end{table}
 
Table \ref{tab:beta_treatments} shows the results of the different usefulness metrics (columns) per treatment (rows).
In the Dyn column, we see that the dynamism treatment yields the highest dynamism score, which confirms our expectation that boosting recency increases intra-list diversity, and hence our implementation is effective.
Serendipity (Ser) and diversity (Div) too see small but significant increases with the dynamism treatment.
The increase in Serendipity may be explained by articles that are boosted by recency, which may take the place of articles that would have better matched user profiles from the \textsc{Baseline} treatment (i.e., recency comes at the cost of personalization). 
For coverage (Cov), there is no significant difference between the two treatments.
Our findings suggest we are successfully adjusting the experimental condition that represents the editorial value under study.

\subsection{Accuracy}
\label{section:online_accuracy}
Now that we've established the dynamism treatment yields more dynamic rankings, and hence the treatment behaves as expected, we set out to answer RQ2. 
Table~\ref{tab:beta_ndcg} shows the accuracy (NDCG) scores macro-averaged over users and days, of the treatments per presentation section. 
Since each section presents a slightly different list of articles to the users, we consider the impact of the dynamism may differ per section. 
However, for none of the sections we observe statistically significant differences between the treatments. 
Paired with the observation of the increased dynamism from Table~\ref{tab:beta_treatments}, we can conclude that we are able to effectively increase dynamism, which represents an important editorial value, without loss of accuracy. 
Finally, we note the discrepancy in accuracy between the offline results from User Study 1 and the online results presented here (NDCG of 0.71 and around 0.5 respectively). 
The observation that offline and online experimental results differ is in line with previous work in the news domain~\cite{garcin2014}.

\begin{table}[t]
    \caption{Average NDCG. 
        * indicates statistically significant differences compared to \textsc{Baseline} with $p$\textless 0.05.}
    \label{tab:beta_ndcg}
    \begin{tabular}{llll}
    \toprule
    & \textsc{MissedLW} & \textsc{MNWidget} & \textsc{MNPage} \\ 
    \midrule
    \textsc{Baseline} & 0.547 & 0.537 & 0.498 \\
    \textsc{Dynamism} & 0.534 & 0.557 & 0.474 \\
    \bottomrule
    \end{tabular}
\end{table}

\subsection{Summary}
In our second user study we find that (i) we can effectively make our news recommender output have more dynamic rankings by boosting recent articles, and (ii) this dynamism treatment does not negatively impact accuracy, suggesting we can incorporate editorial values without hurting accuracy.

\section{Limitations}
\label{section:limitations}

This section describes the limitations of our recommender system design, as well as our experimental setup. 
Since this work was done in the context of a real website with live users, it is not possible to release user data and pre-trained models. 
The model design is not a contribution of the paper, and is itself a replication of the work by \citet{he2014practical}. 
The main paper contributions refer to insights we gained from the data in relation with journalistic values. 
The model features, architecture, and hyperparameters are described in Section~\ref{section:system-design} and provided in the supplementary material~\cite{lu_feng_2020_3758172}, in order to make our experiments replicable.

Our model relies on learning from implicit feedback (i.e., clicks), which brings many challenges, e.g., presentation bias (where clicks are more likely to be observed on top-ranked than lower-ranked items), and negative sampling (where we are only able to observe positive feedback, and have to infer negative)~\cite{4292009}. 
For this study we consider these issues out of scope, and point out that learning from implicit feedback is common in the news domain~\cite{odijkschuth2017,Oard98implicitfeedback,newsrecjannach}. 

The first evaluation in User Study 1 (described in Section~\ref{section:alpha_test}) is limited by the fact that we did not directly evaluate the recommendations through clicks collected on the website, as is common in online evaluations~\cite{shani2011evaluating}. 
Instead, for each ranked list of articles of one user, we pool the clicks from various sources on the website, and perform the evaluation at the level of the ranked list. 
This is because the presentation of the recommendations on the website changed several times in the course of the test, from a separate recommendation page, to a recommendation widget on the front page, to articles with a \textit{recommended for you} label next to them. 
Different displays affected how users received and saw the recommendations, and evaluating at the level of the ranked list makes it possible to combine the diverse recommendation sections from the website.

User Study 2 (Section~\ref{section:beta_test}) is limited by the fact that we only applied a single usefulness treatment (dynamism) out of the four that we studied. 
One reason for this were the real world constraints of calculating usefulness in an online setup. 
As shown in Equation~\ref{eq:usefulness}, our experimental setup incorporates usefulness metrics at the article level, whereas calculating diversity and coverage requires information about all candidate articles, and in the case of inter-list diversity, other users. 
Retrieving and caching multiple confidence scores for different articles and users at the same time adds a significant overhead for the page load time, which was difficult to implement in an online setup. 
Furthermore, as this test was performed with a sizable set of real users, there were concerns about exposing users to too many treatments which might reduce the quality of the recommendations. 
As shown in Table~\ref{tab:beta_ndcg}, even though the NDCG scores are slightly lower for the usefulness treatment in 2 out the 3 sections, the results are not statistically significant. 
This encouraging finding is a good basis for studying the effect of other usefulness treatments in future work.

Similarly, the costly nature of running online tests with actual users on a live website also meant that it was not feasible to claim our test users exclusively for recommender system testing, and our tests were run alongside other tests in parallel. 
These additional tests included various stylistic adjustments of the frontpage (e.g., minor tweaks in font sizes, spacing, etc.), and changes in the order and content of the sections shown on the front page. 
One consequence of this is that both the ways in which the recommendations were displayed, and their surrounding contexts, changed during the tests. 
Different displays will affect how users receive and see the recommendations, and for this reason we resorted to aggregating and averaging the user behavior across presentation modes, as explained in Section~\ref{section:online-test}. 

Finally, a limitation of both studies is the time periods when data was collected -- August for User Study 1, and December for User Study 2, both months that typically exhibit less traffic volume on the FD website, and therefore might not be representative for typical user behavior. The choice of time period was by design, since we wanted to restrict the possible impact of showing users imperfect recommendations. In order to make sure the experiment results are still meaningful, we restricted our opt-in invitations to highly active users, who are more likely to show consistent behavior across periods than infrequent visitors.

\section{Conclusion}

In this paper, we perform two online user studies to better understand how algorithmic recommendation relates to manual curation, and how it steers reading behavior.

In our first user study we compare the output of our recommender system to manually curated editorial lists of articles, and find that recommendations present users with more diversity, se-rendipity, and dynamically changing lists compared to editorially curated lists.
In addition, we compare our users' reading behavior between the month before introducing the recommender system to the month after, and find the more useful recommendations effectively steer our users to more diverse reading behavior, with an overall higher item coverage from the provider's perspective. 

Next, we perform an intervention study where we explicitly incorporate an editorial value that has been deemed important in the context of algorithmic personalization in our recommender system: \emph{dynamism}. 
By incorporating more dynamic recommendations with a re-ranking strategy, we show that we can effectively steer users towards more dynamic reading behavior, without loss of recommendation accuracy. 

Our findings suggest that news recommendation can benefit both news providers and readers.
First, from the provider's perspective an increased overall item coverage means that content will be served to the intended target audiences, which may keep readers more engaged and overall provide economic benefits. 
Second, from the news reader's side, benefits include being served content readers may not have found on a non-personalized, editorially curated front page by themselves, and an increased diversity of news consumption. 
This latter finding, when considered in a broader societal perspective, points towards news recommendation as means of piercing, not creating, filter bubbles. 

In our study we focus on accuracy and usefulness metrics that correspond to short-term behavior. 
Longer term effects of recommendations with increased dynamism on readers' long-term engagement and behavior were out of scope for this study, but could prove beneficial for both providers and readers, and is an aspect worth investigating in future work. 

Finally, our study of incorporating an editorial value without loss of accuracy shows that algorithm design with multiple stakeholders need not be a tradeoff, but can be fruitful for each, as multiple goals can be achieved at the same time.


\begin{acks}
The authors would like to thank Mariella Bastian and Natali Helberger, and the FD Mediagroep AI Team. 
All content represents the opinion of the authors, which is not necessarily shared or endorsed by their respective employers (past or present).
\end{acks}

\bibliographystyle{ACM-Reference-Format}
\balance
\bibliography{reference}


\end{document}